\documentclass[twocolumn]{aastex62}

\graphicspath{{./}{figures/}}

\received{May 13}
\revised{July 23}
\accepted{\today}
\submitjournal{ApJ}

\usepackage{color}
\shorttitle{Sample article}
\shortauthors{Mori et al.}

\begin{document}
\newcommand{\kataoka}[1]{\textcolor{MyDarkGreen}{[Kataoka: #1]}}

\title{An Observational Study for Grain Dynamics in the AS 209 Disk with Submillimeter Polarization \footnote{Released on January, 8th, 2018}}

\author[0000-0002-0786-7307]{Tomohiro Mori}
\affil{The Institute of Astronomy, the University of Tokyo, 2-21-1 Osawa, Mitaka, Tokyo 181-8588, Japan}

\author{Akimasa Kataoka}
\affiliation{National Astronomical Observatory of Japan, 2-21-1 Osawa, Mitaka, Tokyo 181-8588, Japan}

\author{Satoshi Ohashi}
\affiliation{RIKEN Cluster of Pioneering Research, 2-1, Hirosawa, Wako-shi, Saitama 351-0198, Japan}

\author{Munetake Momose}
\affiliation{College of Science, Ibaraki University, 2-1-1 Bunkyo, Mito, Ibaraki 310-8512, Japan}

\author{Takayuki Muto}
\affiliation{Division of Liberal Arts, Kogakuin University, 1-24-2 Nishi-Shinjuku, Shinjuku-ku, Tokyo 163-8677, Japan}

\author{Hiroshi Nagai}
\affiliation{National Astronomical Observatory of Japan, 2-21-1 Osawa, Mitaka, Tokyo 181-8588, Japan}

\author{Takashi Tsukagoshi}
\affiliation{National Astronomical Observatory of Japan, 2-21-1 Osawa, Mitaka, Tokyo 181-8588, Japan}



\begin{abstract}
We present 870 $\micron$ ALMA polarization observation toward the Class I\hspace{-.1em}I protoplanetary disk around AS 209, which has concentric, multiple gaps and rings. We successfully detect the polarized emission and find that the polarization orientations and fractions have distinct characteristics between the inner and outer regions. In the inner region, the polarization orientations are parallel to the minor axis of the disk, which is consistent with the self-scattering model. 
The mean polarization fraction in the region is $\sim$0.2\%, which is lower than the expected value when the grains have the maximum polarization efficiency, which corresponds to $\lambda$/2$\pi$ $\sim$ 140 $\micron$ in grain radius. In the outer region, we detect $\sim$1.0\% polarization and
find that the polarization orientations are almost in the azimuthal directions. Moreover, the polarization orientations have systematic angular deviations from the azimuthal directions with $\Delta$$\theta$ $\sim$ 4\fdg5 $\pm$ 1\fdg6. The pattern is consistent with a model that radially drifting dust grains are aligned by the gas flow against the dust grains. We consider possible scenarios of the grain dynamics at the AS 209 ring which can reproduce the polarization pattern. However, 
the directions of the observed angular deviations are opposite to what is predicted under the fact that the disk rotates clockwise. This poses a question in our understandings of the alignment processes and/or grain dynamics in protoplanetary disks.
\end{abstract}



\section{Introduction} \label{sec:intro}
In protoplanetary disks, submicron-sized dust grains coagulate to form larger aggregates, and eventually to form planets.
However, there are several long-standing questions which make it difficult to realize the coagulation.
One of the most critical obstacles is the radial drift barrier. This is a problem that dust grains rapidly drift toward stars due to a headwind from gas, which rotates slower than the dust grains \citep{adachi1976, weidenschilling1977, nakagawa1986, takeuchi2002, takeuchi2005, brauer2007, brauer2008}.

One of the promising scenarios to avoid the barrier is trapping dust grains at local pressure maxima that can stop the inward drift, and potentially leading to efficient coagulation of the dust grains \citep{whipple1972, pinilla2012}.
In fact, high spatial resolution dust continuum observations by using Atacama Large Millimeter Array (ALMA) have revealed that many disks have a series of the concentric rings, which is possibly the result of the grain trapping at the gas pressure maxima \citep{alma2015, isella2016, tsukagoshi2016, loomis2017, andrews2016, andrews2018, fedele2018, guzman2018, long2018}. 
The efficiency of the grain trapping is determined by how well dust grains couple to gas, which is dependent on the gas surface density and the grain size.
Thus, observational constraints on the grain size are essential to discuss the grain radial drift and trapping.

Millimeter-wave polarization observations provide a unique opportunity to constrain the grain size.
The millimeter-wave polarization of disks is believed to originate from a combination of grain alignment and self-scattering.
Previously proposed sources which can align grains in disks are magnetic fields, radiative gradients, and gas flow directed on grains \citep{cho2007, tazaki2017, gold1952, lazarian2007, kataoka2019}. The thermal emission from elongated dust grains aligned with the sources can be observed as polarized emissions. 
The other mechanism, self-scattering, is scattering-induced polarization where dust grains of sizes comparable to the wavelengths scatter the thermal dust emissions. When the dust thermal emissions are anisotropic in disks, the scattered emissions can be observed as polarized emissions \citep{kataoka2015}.

The first detection of the millimeter-wave polarization in a protoplanetary disk was reported in the HL Tau disk with Sub Millimeter Arrays (SMA) and Combined Array for Research in Millimeter-wave (CARMA) \citep{stephens2014}. Subsequently, in the ALMA era, the number of detection of millimeter-wave polarization has been increasing owing to the high-sensitivity and high spatial resolution observations \citep{bacciotti2018, cox2018, girart2018, harris2018, hull2018, lee2018, ohashi2018, sadavoy2018}. These observations revealed that polarized emissions due to self-scattering ubiquitously appear, implying a presence of 100 micron-sized grains in the many disks. Grain alignment also has been considered to contribute to the observed polarization in some disks \citep{kataoka2017, stephens2017, ohashi2018}.
In this work, we perform 870 $\mu$m ALMA polarization observations of the disk around AS 209 and discuss the origins of the polarization.

AS 209 is a Classical T Tauri star in the Ophiuchus star forming region at a distance of 121 $\pm$ 2 pc \citep{gaia2018}.
It has a mass of 0.9 $M_{\odot}$ \citep{bouvier1992}, spectral type of K5 \citep{luhman1999} and a luminosity of 1.5 $L_{\odot}$ \citep{natta2006}. 
The disk mass was estimated to 0.028 $M_{\odot}$ with SMA continuum observation assuming 100:1 dust-to-gas mass ratio \citep{andrews2009}. More recently, \citet{favre2019} conducted CO isotopologue observation with ALMA and derived the gass mass of 3 $\times$ $10^{-3}$ $M_{\odot}$, which is less massive compared to the previous study \citep{andrews2009}.
ALMA 1.3 mm dust continuum observations revealed that the disk has two prominent rings at 74 and 120 au and at least three faint rings in the inner 60 au \citep{fedele2018, guzman2018}. 
Planet-disk interaction is one promising gap opening mechanism. Recent studies using 3D hydrodynamical simulations found that the ringed structure may be induced by torque from a Saturn-like mass planet at $\sim$100 au \citep{fedele2018, zhang2018, favre2019}.
The radial profile of the grain size of the AS 209 disk is constrained by measuring spectral index \citep{perez2012, tazzari2016}. They found that the radial profile of the opacity index $\beta$ radially increases, suggesting that the grain size radially decreases from $\sim$2 cm to $\sim$0.2 cm.

This paper is organized as follows.
In Section 2, we briefly summarize observation and data reduction processes. In Section 3, we describe characteristics of the observed polarized emission based on the polarization morphology and fraction. In Section 4, we explore the origin of the polarized emission and discuss a grain property and dynamics in the AS 209 disk. 
In Section 5, we conclude this paper.

\section{Observations} \label{sec:observation}
The continuum polarization observations at 870 $\mu$m were carried out on 2018 May 16 during ALMA Cycle 5 operation. The antenna configurations were C43-2 with 43 antennas. 
In total four spectral windows (spws), two for lower sideband, another two for upper sideband, were taken in time division mode (TDM).  Those four spws were centered at the central frequencies of 336.5, 338.4, 348.5 and 350 GHz. 
The effective bandwidth of each spw is 1.875 GHz, providing the bandwidth of $\sim$7.5 GHz. The bandpass, complex gain, and polarization calibrators were J1751-0939, J1733-1304 and J1924-2914, respectively. The polarization calibrator was observed in four different scans with a scan length of $\sim$8 minutes at different hour angles to cover a wide range of parallactic angle.
The total integration times for the target were 30.89 minutes in the observation.

The reduction and calibration of the data were done with the CASA version 5.1.1 \citep{mcmullin2007}.
We follow the data reduction process given by \citet{nagai2016}. For the imaging, we perform the interactive CLEAN deconvolution by using the CASA task {\tt tclean}. We employ Briggs weighting with a robust parameter of 0.5. The beam size in the final product is 0.\arcsec94 $\times$ 0.\arcsec62 which corresponds to $\sim$114 $\times$ 75 au at the distance of 121 pc.

With the obtained Stokes $Q$ and $U$, we derive the polarized intensity ($PI$). Note that $PI$ has a positive bias because it is always positive value even if the Stokes $Q$ and $U$ have negative values. This bias has a non-negligible effect in low-signal-to-noise observations. We thus derive the debiased polarized intensity with the following equation presented in \citet{vaillancourt2006} and \citet{hull2015}, 
\begin{equation}
PI = \sqrt{Q^2 + U^2 - \sigma_{PI}^2}
\end{equation}
where $\sigma_{PI}$ is an error of the polarized intensity, which is derived to be $\sigma_{PI}$ = 2.7 $\times$ $10^{-5}$ Jy beam$^{-1}$ with the error propagation of Stokes $Q$ and $U$. 

Polarization fraction (PF) and polarization angle ($\theta$) are also derived with the following equation.
\begin{equation}
PF = PI/I
\end{equation}
\begin{equation}
\theta = \frac{1}{2}\arctan{\frac{U}{Q}}
\end{equation}
The 1$\sigma$ error of the polarization angle $\sigma_{\theta}$ is calculated in each pixel in the image with a following equation.
\begin{equation}
\sigma_{\theta} (^\circ) = 0.5 \times 180/\pi \times \sqrt{(U \times \sigma_{Q})^2 + (Q \times \sigma_{U})^2}/PI^2
\end{equation}
where $\sigma_{Q}$ and $\sigma_{U}$ are the rms noise of Stokes $Q$ and $U$, respectively. In addition, instrumental angle errors can systematically contribute to the uncertainties of the angles. This systematic error ($\Delta\chi$) is $\Delta\chi$ $\sim$ 2$^\circ$/$\sqrt{N}$, where $N$ is the number of antennas \citep{nagai2016} and thus $\Delta\chi$ $\sim$ 2$^\circ$/$\sqrt{43}$ = 0\fdg3 in this observation.
All of these values presented above are derived only where the detection is above the threshold 3$\sigma_{PI}$.

\section{Results} \label{sec:floats}
\begin{figure*}[h!]
\centering
\includegraphics[height=15cm]{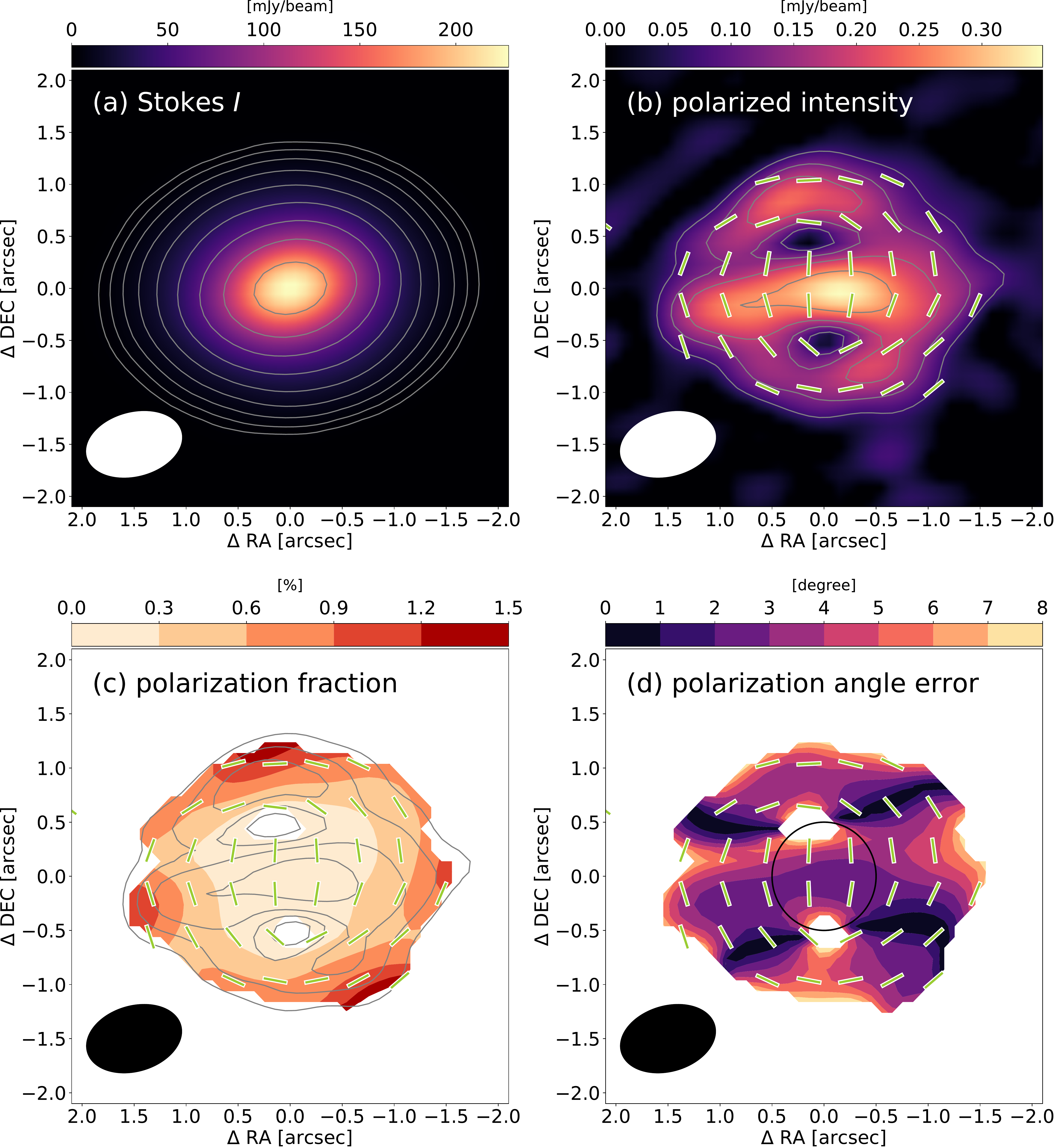}
\caption{Upper left (a): The total intensity (Stokes $I$) of the continuum emission at 870 $\mu$m. The solid contours represent total intensity with levels of 100$\--$10000 $\times$ $\sigma$$_I$ (=60 $\mu$Jy beam$^{-1}$) in log space. 
The beam with the size of $0.\arcsec94$ $\times$ $0.\arcsec62$ and position angle of $-$75\fdg3 is shown in the bottom left with the white ellipse.
Upper right (b): Polarized intensity on a linear scale. The solid contour levels are (3, 5, 7, 10) $\times$ $\sigma$$_{PI}$ (= 27 $\mu$Jy beam$^{-1}$). The polarization vectors are presented where polarized intensity is larger than 3$\sigma$$_{PI}$. We set the length of the polarization vectors to be the same.
Lower left (c): The polarization fraction overlaid with the vectors. The solid contours show polarized intensity as with the $PI$ map. The polarization fraction where polarized intensity is less than 3$\sigma$$_{PI}$ is removed. The synthesized beam is also presented with the black ellipse.
Lower right (d): The color map of the 1$\sigma$ polarization angle error ($\sigma_{\theta}$). The synthesized beam and polarization vectors are also overlaid. The overlaid circle at the center represents the boundary of the inner and outer regions with the radius of 0.\arcsec5.}
\label{fig:main_results}
\end{figure*}

The main results of the observation are summarized in Figure \ref{fig:main_results}.
Figure \ref{fig:main_results}(a) shows the total intensity (Stokes $I$) of the protoplanetary disk around AS 209 with the color scale and contours. 
The substructures discovered in the high-resolution observations \citep{fedele2018, guzman2018} are not identified in our Stokes $I$ image because the beam size 0.\arcsec94 $\times$ 0.\arcsec62 is not small enough to resolve the rings or gaps with the size of $\sim$$0.\arcsec04$ \citep{guzman2018}. The integrated flux density is 489 $\pm$ 50 mJy, which is consistent with the previously obtained value, 577 $\pm$ 58 mJy within the uncertainty \citep{andrews2009}.

Figures 1(b) and 1(c) show the polarized intensity and polarization fraction, respectively. The polarization vectors are also overlaid where $PI$ is larger than 3$\sigma_{PI}$ = 81 $\mu$Jy beam$^{-1}$. The vectors are plotted twice per synthesized beam in each direction based on Nyquist sampling.
Figure \ref{fig:main_results}(d) shows the map of $\sigma_{\theta}$ in degrees estimated with the equation (4). 
The typical 1$\sigma$ polarization angle error is $\sigma_{\theta}$ $\sim$ 1$\--$$4^{\circ}$. Note that the systematic angle error ($\Delta\chi$ = 0\fdg3) explained in the former section is not included in Figure \ref{fig:main_results}(d).

\begin{figure}
\centering
\includegraphics[height=8cm]{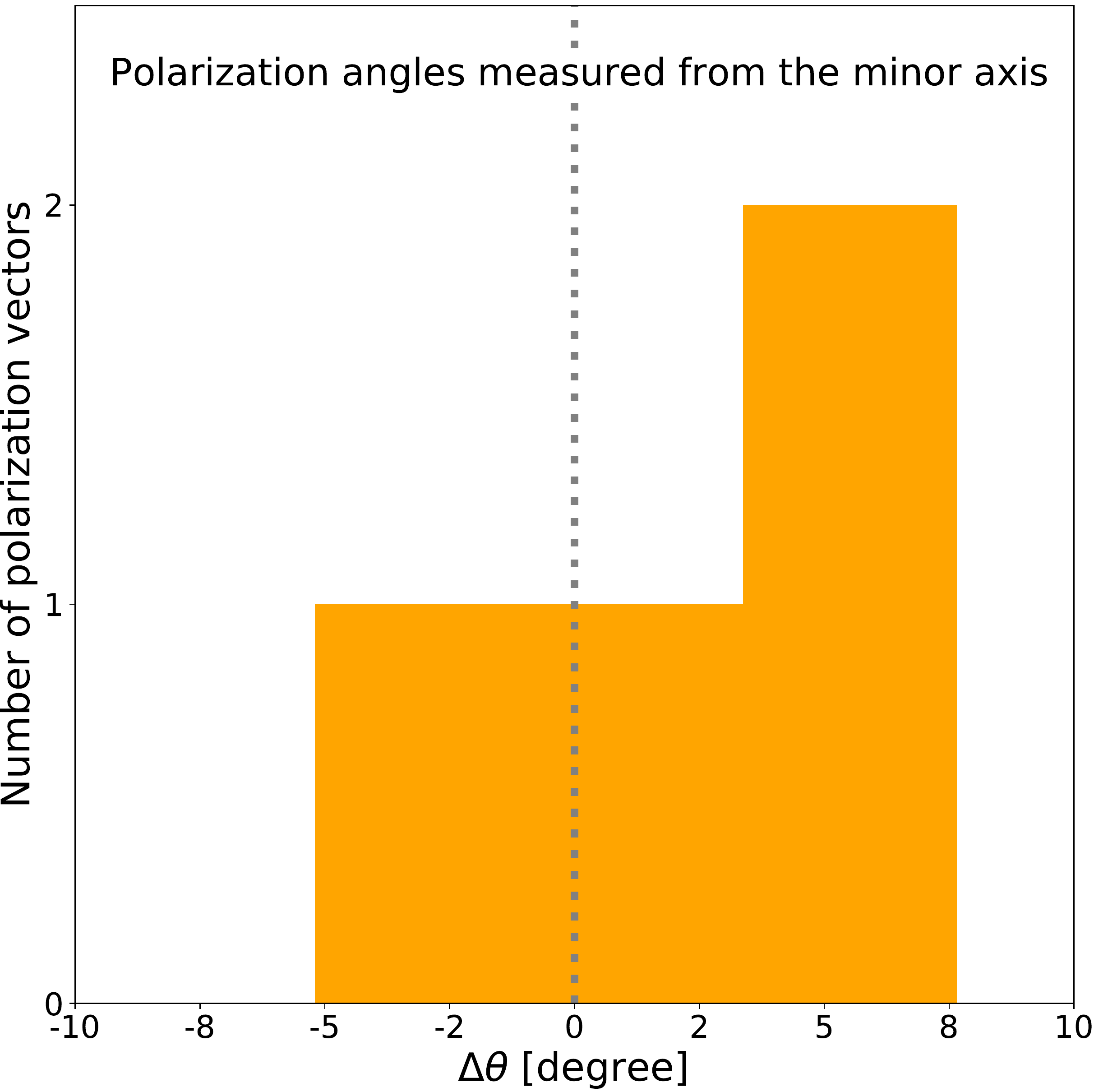}
\caption{
The distribution of the polarization angles measured from the minor axis in the inner 0.\arcsec5 region. Three bins are presented with the width of $\sim$4$^\circ$.
The gray dotted line represents the positions of the minor axis PA.}
\label{fig:inner_histogram}
\end{figure}

The polarization orientations show distinct patterns between the inner and outer regions. In the inner 0.\arcsec5 region, which is illustrated by the circle in Figure \ref{fig:main_results}(d), the polarization orientations seem to be parallel to the minor axis.
In Figure \ref{fig:inner_histogram}, we plot a histogram of the deviations of the polarization angles from the minor axis to see if the vectors are completely aligned with the minor axis. 
We use the major axis position angle (PA) of 85\fdg76 and minor axis PA of $-$4\fdg24 with reference to \citet{guzman2018}. 
Due to the large beam size, only four vectors are plotted with the bin width of $\sim$4$^\circ$. The plotted angular deviations ($\Delta\theta_{minor}$) are $\Delta\theta_{minor}$ = 6\fdg7 $\pm$ 2\fdg8, $-$5\fdg1 $\pm$ 2\fdg5, 1\fdg4 $\pm$ 4\fdg3, and 7\fdg7 $\pm$ 3\fdg6. Note that the histogram with the small sample size is perhaps not helpful to visualize the angle distribution. However, such analysis will be helpful for future observations with smaller beam size, and thus we keep the histogram in this paper.

The center of the histogram seems to be around $\sim$0$^\circ$, but three of the $\Delta\theta_{minor}$ values significantly deviate from 0. 
Given the fact that the sample size is small and each vector is not necessarily independently sampled, we cannot conclude whether the spread is real or not. One possibility is that the spread comes from beam dilution, where polarization in the outer regions can cause some angular deviations from the minor axis.
Although an observation with a smaller beam size is necessary to confirm the statement, we conclude that the polarization orientations in the inner region are consistent with the direction of the minor axis.

In the outer $0.\arcsec5$ region, on the other hand, the polarization orientations seem to be aligned in the azimuthal directions. To interpret the polarization, we proceed to a further analysis of the polarization angles. 
First, we discuss whether the observed polarization orientations are consistent with circular or elliptical patterns.
The circular pattern is a concentric circle on the image plane, and the elliptical pattern is a trajectory that traces the same orbits in the inclined disk. We derive the tangents of the circle and the ellipse at each location of the polarization vectors. 
For the derivations of the elliptical tangents, we use the same PA of the major axis as the previous discussion and the inclination of $34\fdg88$ \citep{guzman2018}.
Figures 3(a) and (b) depict comparisons between the observed polarization orientations and the circular (a) or elliptical (b) tangents overlaid on the Stokes $I$ map.
Figures 3(c) and (d) show the distribution of the angular deviations from the circular (c) and elliptical (d) tangents, 
each of which is overlaid with the best-fitted Gaussian function.
\begin{figure*}[h!]
\centering
\includegraphics[height=15cm]{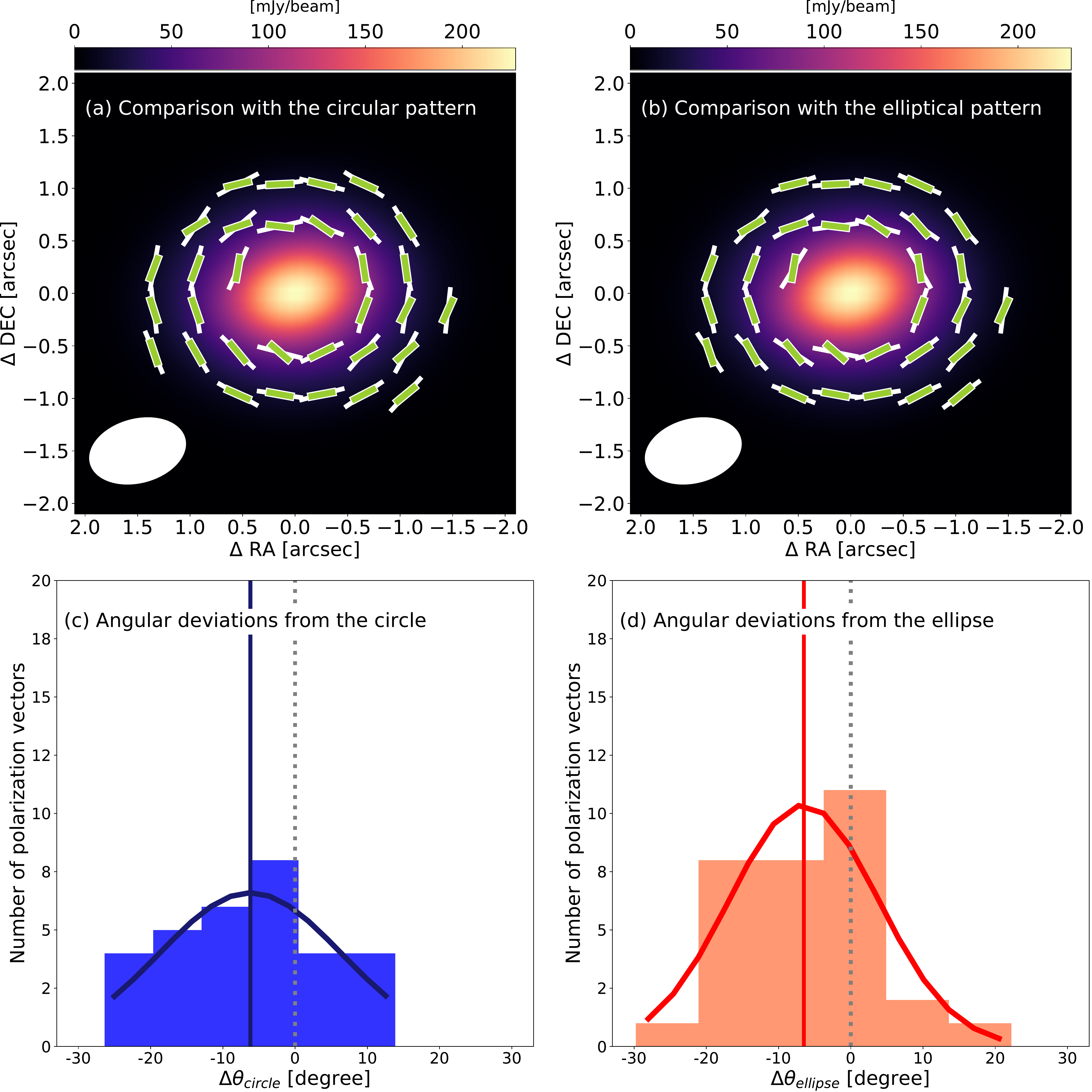}
\caption{(a) Comparison between the polarization vectors and circular tangents in the outer 0.\arcsec5 region. The green and white lines represent the polarization vectors and circular tangents, respectively.
(b) Comparison between the polarization vectors and elliptical tangents. The green and white lines are represented as with (a). 
(c) The histogram of the angular differences between the polarization vectors and the circular tangents, both of which are presented in (a). The best-fitted Gaussian is overlaid to the histogram. The blue straight line represents the center of the distribution. The gray dotted line represents the $\Delta$$\theta_{circle}$ = 0$^\circ$ position.
(d) The histogram of the angular differences between the polarization vectors and the ellipse, both of which are presented (b). The best-fitted Gaussian and the center of the distribution are overlaid as with (c) with red lines. $\Delta$$\theta_{ellipse}$ = 0$^\circ$ position is also presented as with (c).
}
\label{fig:outer_histogram}
\end{figure*}

\begin{figure}[h!]
\centering
\includegraphics[width=7cm]{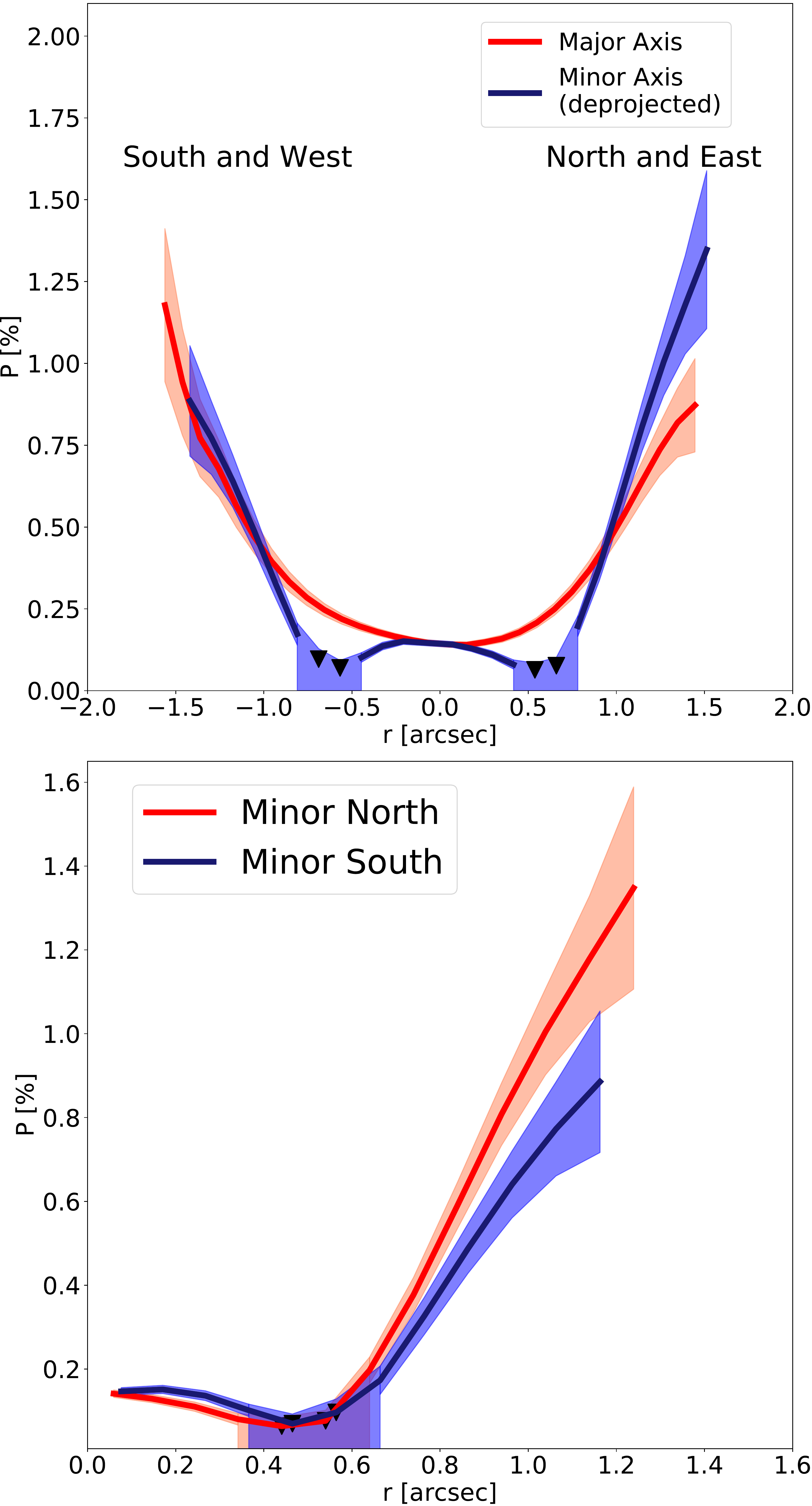}
\caption{
Upper panel: radial profiles of the polarization fraction along with the major (red) and minor (blue) axes. To correct the disk inclination ($i$ = $34\fdg88$), the minor axis profile is enlarged by multiplying the radius by 1/$\cos$ $i$ for the deprojection. The shade in each profile represents the 1$\sigma$ error regions of the polarization fraction. The upper limits and shade are also presented in the region where the polarized intensity is less than 3$\sigma_{PI}$.
Bottom panel: minor axis profiles of the polarization fraction in the north (red) and south (blue), both of which are extracted from the upper panel. The deprojection is not applied to the profiles.}
\label{fig:polarization_fraction}
\end{figure}

The histogram of the angular deviations from the elliptical tangents shows a little narrower width than that from the circular tangents. 
We confirm this by performing the Gaussian fittings of the histogram, showing that the standard deviations are $\sim$10$^\circ$ and $\sim$11$^\circ$ in the elliptical and circular case, respectively.
Therefore, the polarization pattern is more consistent with the elliptical pattern rather than the circular pattern. We also conduct a $\chi^2$ test to statistically examine the above discussion and conclude that the elliptical pattern is preferred indeed (see Appendix A).

Then, we investigate if there is a systematic angular difference between the polarization angles and the elliptical tangents. If the polarization vectors are completely aligned with the ellipse, 
the mean of the angular differences $\Delta$$\theta_{ellipse}$ would be $0^\circ$. However, the histogram of the angular differences shows a certain shift from $\Delta$$\theta_{ellipse}$ = $0^\circ$. 
We calculate a weighted mean of $\Delta$$\theta_{ellipse}$ with the $\Delta$$\theta_{ellipse}$ and $\sigma_{\theta}$ maps (Appendix A). This results in the weighted mean of $\Delta$$\theta_{ellipse}$ = $-$$4\fdg5$ $\pm$ $1\fdg6$, which is significantly shifted from $0^\circ$. 
This suggests that the polarization orientations have certain angular deviations from the elliptical tangents with almost the same degrees and directions. Note that a beam dilution due to the large and flattened beam can artificially generate the angular deviation from the azimuthal directions. Although we need detailed modeling to examine the effect, the obtained orientations which systematically deviate from the azimuth cannot be caused only by the beam dilution. 

The polarization fractions are another clue to interpreting the polarization.
To compare the polarization fractions between the major and minor axes, we plot radial profiles of the polarization fractions along the major and deprojected minor axes in Figure \ref{fig:polarization_fraction} (upper panel).
The negative values in the horizontal axis represent the south and west regions, while the positive values represent the north and east regions.

As Figures 1(c) and Figure \ref{fig:polarization_fraction} show, polarization fractions are not the same between the two regions as well as the polarization orientations.
In the inner 0.\arcsec5 region, the polarization fractions are almost uniform at $\sim$0.20$\%$.
Note that the rms noise of the polarization fraction is $\sim$0.01$\%$ and thus the 0.20$\%$ polarization fraction is significant.
In the outer 0.\arcsec5 region, on the other hand, the polarization fractions gradually increase from $\sim$0.2$\%$ to $\sim$1.0$\%$ with increasing radius. 
The gradual increase is likely due to the beam dilution between the inner and outer region, each of which shows distinct polarization orientations and thus cancel out each other.
This can reduce the polarization fractions 
near the $0.\arcsec5$ regions where the distinct polarization orientations coexist in the inner and outer regions.
The effect becomes weaker with increasing the distance from the center.
Thus, the typical polarization fraction in the outer region is likely $\sim$1.0\%, which is obtained in the outermost region.

Another feature in the polarization fractions is that the radial profile along the minor axis shows a certain asymmetry that polarization fractions increase more sharply in the north region. The bottom panel of Figure \ref{fig:polarization_fraction} shows the minor axis profiles extracted from the upper panel. Indeed, the polarization fractions in the north region reach higher values of $\sim$1.2\%. The asymmetry is probably not caused by the beam dilution since its effect should be the same between the north and south regions. However, the difference of the profiles is only 1$\sigma$, and moreover the detection of the polarized intensity is somewhat marginal with the signal-to-noise ratio of $\sim$3$\--$5$\sigma_{PI}$.
Therefore, although the asymmetry may reflect some physical origins, we cannot robustly conclude if it is real with this data.

\section{Discussion} 
We detect the $\sim$0.20\% and $\sim$1.0\% polarization in the inner and outer regions, respectively. 
The polarization orientations are parallel to the minor axis in the inner region, while they are in the azimuthal (elliptical) directions in the outer region.
These distinct characteristics of the polarized emission imply the distinct origins of the polarized emission between the regions. First, we describe the possible origins of the polarization and predicted polarization pattern in subsection 4.1.
Then, we explore the origins in the AS 209 disk and possible models for the grain properties and grain dynamics in the AS 209 disk in the inner and outer regions in subsection 4.2. and 4.3.

\subsection{The polarization pattern in the different theories}
The possible origins of the millimeter-wave polarization from protoplanetary disks are the grain alignment or self-scattering. The grain alignment models include magnetic, radiative and mechanical alignment, where dust grains are aligned with the magnetic fields, radiative gradient and gas flow, respectively.
We quickly review the currently proposed scenarios of the millimeter-wave polarization and summarize the phenomenological differences.

Long axes of the magnetically aligned dust grains are perpendicular to magnetic fields \citep{davis1951, cho2007}. The toroidal magnetic fields are thought to be amplified with magnetorotational instability (MRI) in disks \citep{brandenburg1995, fromang2006, bai2013, suzuki2014}.
Thus, the resultant polarization would show a radial pattern in face-on disks \citep{brandenburg1995}.

Grain alignment with radiation fields makes grain long axes perpendicular to radiation gradients \citep{cho2007, tazaki2017}. Thus, the outgoing radiation gradients would produce a circular pattern in polarization vectors in face-on disks. We note that \citet{yang2019} pointed out that the polarization orientations remain circular patterns even in inclined disks.

\begin{deluxetable*}{cccr}
\tablecaption{The origin and expected polarization morphology (inclined disks)\label{tab:mathmode}}
\tablecolumns{2}
\tablenum{1}
\tablewidth{0pt}
\tablehead{
\colhead{Origin} &
\colhead{Polarization morphology} &
}
\startdata
magnetic alignment & radial (if toroidal) \\
radiative alignment & circular \\
mechanical alignment (Gold mechanism) & elliptical \\
mechanical alignment (helicity-induced, small grain) & circular (if $\delta$$v_r$ $\gg$ $\delta$$v_{\phi}$) \\
mechanical alignment (helicity-induced, large grain) & spiral-like (if $\delta$$v_r$ $\sim$ $\delta$$v_{\phi}$) \\
self-scattering & minor axis \\
\enddata
\end{deluxetable*}

Grain alignment with the ambient gas flows, which is called mechanical alignment, makes grain long axes to be either parallel or perpendicular to the gas velocity against the dust grains. The alignment parallel to the gas 
occurs for subsonic gas flows, which is called Gold mechanism \citep{gold1952}. The aligned dust grains would produce elliptical polarization pattern in inclined disks  \citep{yang2019}. Dust grains can also be aligned perpendicular to the gas flow onto the dust grains when they have certain helicity \citep{lazarian2007}. The resultant polarization is perpendicular to the gas velocity against the dust grains. This helicity-induced grain alignment occurs for subsonic gas. Since gas velocity against dust grains is subsonic, helicity-induced alignment likely occurs in protoplanetary disks \citep{kataoka2019}.

The polarization orientations for mechanical alignment are determined by gas velocity against dust grains. We express the radial and azimuthal components of the gas velocity against the dust grains with $\delta$$v_r$ and $\delta$$v_{\phi}$, respectively.
The gas velocity against the dust grains is also determined by how well the dust grains are coupled to gas, which is denoted with the Stokes number (St). The Stokes number is the dust stopping time normalized with the Keplerian timescale and is given by St = $\frac{\pi}{2}$$\frac{a\rho_s}{\Sigma_g}$, where $a$ is the grain size, $\rho_s$ is the internal density of the dust grains and $\Sigma_g$ is the gas surface density \citep{birnstiel2009}. 
We briefly describe the expected polarization patterns for the helicity-alignment model in \citet{kataoka2019}, assuming that the dust grains radially drift due to the headwind from the gas.

When the Stokes number is much smaller than unity, which corresponds to small grain size, the headwind of the gas on the dust grain is dominated by the radial component ($\delta$$v_r$ $\gg$ $\delta$$v_{\phi}$). 
The polarization orientations are perpendicular to the gas velocity against the dust grains, and thus the resultant polarization pattern is circular.
When the Stokes number is close to unity, the larger dust grains are decoupled from the gas and radially drift due to the gas headwind, resulting in the comparable velocity fields against the dust grains between the radial and azimuthal components ($\delta$$v_{\phi}$ $\lesssim$ $\delta$$v_r$).
As a consequence, the synthetic relative velocity is inclined with respect to the azimuthal directions, and thus the polarization orientations show a spiral-like pattern. The deviation from the circular direction at each location on the sky would be arctan($\delta$$v_r$/$\delta$$v_{\phi}$) \citep{kataoka2019}.

The polarization orientations for the scattering-induced polarization are determined by incoming flux distributions around the dust grains \citep{kataoka2015}. In the inclined disk such as the AS 209 disk, the polarization orientations are parallel to the minor axis because the flux coming parallel to the major axis is generally stronger than that parallel to the minor axis \citep{pohl2016, kataoka2017}. \citet{yang2017} pointed out that the polarization patterns can be modified by the spatial distributions of the optical depth.
When the disk is optically thin in the outer region, the polarization orientations are perpendicular to the radiative gradient in the disks. The outgoing radiative gradient leads to the azimuthal pattern in the outer region. However, the model assumed smooth surface density structures, which has no rings and gap. In the case of the AS 209 disk, the radiative gradient of the flux distribution produced by the ringed structure would be too small to produce the azimuthal pattern. Thus, if the scattering dominates the observed polarized emission, the polarization orientations are parallel to the minor axis even if the disk is optically thin in the outer region.

In Table 1, we summarize the expected polarization patterns explained above.
In the following subsections, based primarily on the polarization patterns shown in Table 1,
we explore the origin of the polarization separately in the inner and outer regions.

\subsection{The origin of the polarization in the inner region}
In the inner region, the polarization vectors are aligned with the minor axis (Figure \ref{fig:inner_histogram}). Although certain deviations from the minor axis exist, the observed orientations which are parallel to the minor axis are consistent with the self-scattering model in the inclined disks (Table 1).
Therefore, we conclude that self-scattering dominates the polarized emission in the inner region of the AS 209 disk at 0.87 mm.

The polarization fraction in the inner region is $\sim$0.2 $\%$ (Figure \ref{fig:main_results}(a) and 4).
The self-scattering model predicts the polarization fraction of 2$\--$3$\%$, 
when the grain population has a single power law with the maximum grain size of $a_{max}$ $\sim$ $\lambda$/2$\pi$ $\sim$ 140 $\mu$m \citep{kataoka2015, kataoka2016a}. Polarization fractions obtained in the previous observations for the inclined disks were lower than 2$\--$3$\%$, but significantly higher than that in this observation.
For example, the IM Lup and HL Tau disks show scattering-induced polarization with the fractions of $\sim$1.2\% and $\sim$0.6\% in the central regions \citep{hull2018, stephens2017}. Relatively low polarization fraction in the AS 209 disk can be explained by two possibilities in the grain size populations. One is that the grain population has a single power law with a maximum grain size of $a_{max}$, which is a few times larger or smaller than $\sim$140 $\mu$m \citep{kataoka2016a}. If this grain model is correct, $a_{max}$ is roughly estimated to be $\sim$50 $\micron$ or $\sim$400 $\micron$. The other possibility is that two size population model where $a_{max}$ $\sim$ 140 $\mu$m in one population but $a_{max}$ in the other population is significantly smaller or larger than 140 $\mu$m.
This additional grain population contributes to unpolarized Stokes $I$ emission but not to polarized emission, and thus reduces polarization fractions. 

The maximum grain size has been constrained by \citet{perez2012} and \citet{tazzari2016} with the analysis of the spectral index.
They found that the spectral index radially increase with the range $\beta$ = 0.5$\--$1, and
yielded the grain size of $a_{max}$ $\sim$ 0.5$\--$2 cm in inner 60 au.
These grain sizes are much larger than that from the polarization.

The size discrepancy may come from the contamination of the optically thick emission in low-resolution observations. Although measurements of the grain size assume the thermal emission is totally optically thin in millimeter wavelength, \citet{tripathi2017} pointed out that the assumption is not necessarily correct based on a large sample of low-resolution disk images. However, \citet{perez2012} and \citet{tazzari2016} revealed that the millimeter continuum emission from the AS 209 disk is optically thin at all radii, suggesting that the discrepancy of the grain sizes are not due to the optically thick emission.

The assumption on the grain composition also strongly affects the spectral indices values, leading to grain size uncertainties.
For example, \citet{testi2014} showed that compact grains which composed of silicate, carbonaceous and water ice take $a_{max}$ $\sim$ 0.5$\--$5 cm in $\beta$ = 0.5$\--$1 range while compact grains which composed only of silicate and carbonaceous material take $a_{max}$ $\sim$ 0.5$\--$5 mm in the same $\beta$ = 0.5$\--$1 range (see Figure \ref{fig:polarization_fraction} of \citet{testi2014}). 
The one size population model with $a_{max}$ = 500 $\mu$m and the latter grain composition model described above is one of the solutions to reconcile the two studies. 

More recent studies pointed out that the 
observed low spectral indices can be reproduced when the effect of scattering is included \citep{baobab2019, zhu2019}. This means that measurement of the grain size ignoring the opacity of scattering can lead to overestimating of the grain size. Thus, including the effects of scattering into continuum modeling is needed to reconcile the discrepancy of the grain properties between the grain size measurements.

\subsection{The origin of the polarization in the outer region}
In the outer 0.\arcsec5 region, the polarization orientations seem to be in the azimuthal directions. By conducting the detailed analysis on the orientations, we find that the orientations are consistent with the elliptical pattern rather than the circular pattern and moreover the vectors have the systematic angular deviations from the elliptical tangents with the mean value of $\Delta\theta_{ellipse}$ = $-$4\fdg5 $\pm$ 1\fdg6. The spiral-like pattern can be produced only with the mechanical alignment model (Table 1). Therefore, the observed polarization likely originates from the dust grains which are aligned by the gas flow against the dust grains. 

\begin{figure}
\centering
\includegraphics[height=7cm]{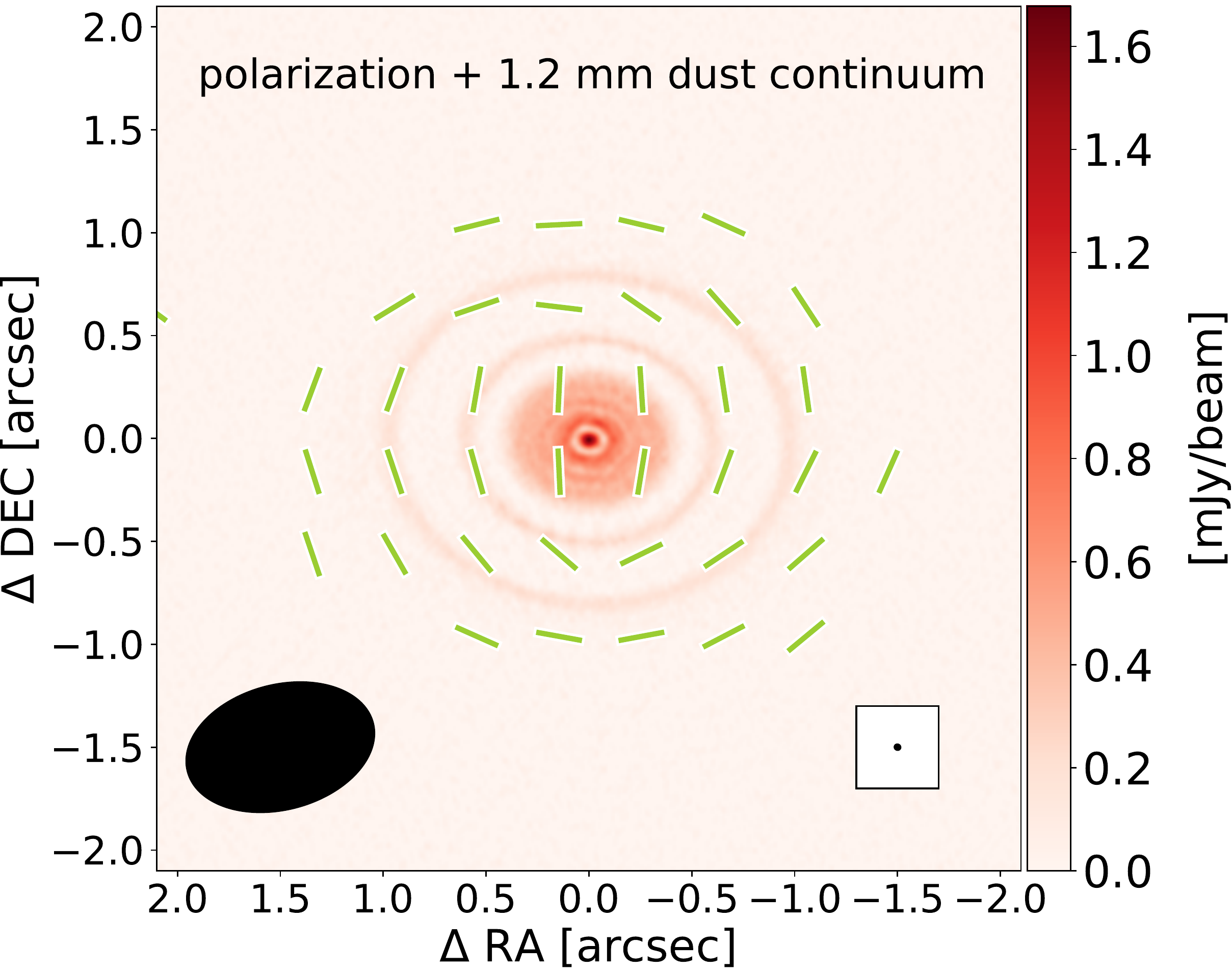}
\caption{The polarization vectors and the high resolution image obtained in the DSHARP program \citep{guzman2018}.
The synthesized beam size in our and DSHARP observations are presented in the lower left and lower right, respectively.}
\label{fig:dsharp_vector}
\end{figure}

To interpret the polarization, we explore where the polarized emission comes from. However, the large beam size prevents us from exploring the emitting regions only with the polarization data. Thus, in Figure \ref{fig:dsharp_vector}, we compare the polarization vectors with the previous higher resolution observation by \citet{guzman2018}.
Although it is difficult to link the positions of the vectors to that of high resolution image due to the large beam size gap, roughly speaking, the polarization in the outermost region at $\sim$$1.\arcsec0$ likely comes from the outermost ring at 120 au in the higher resolution image.
Thus, the emitter of the polarization is likely the dust grains at the 120 au ring.

Since the polarization pattern is presumably related to the grain dynamics in the disk, we consider possible grain dynamics to reproduce the spiral-like pattern.
\citet{kataoka2019} has already considered the relationships between the grain dynamics and polarization patterns.
However, we cannot directly apply the model to the observed pattern because the assumed disk in the model has a smooth surface density profile which is quite different from that of the AS 209 disk. Therefore, we qualitatively discuss the polarization pattern when mechanical alignment occurs at the ring.

\begin{figure}
\centering
\includegraphics[height=4.5cm]{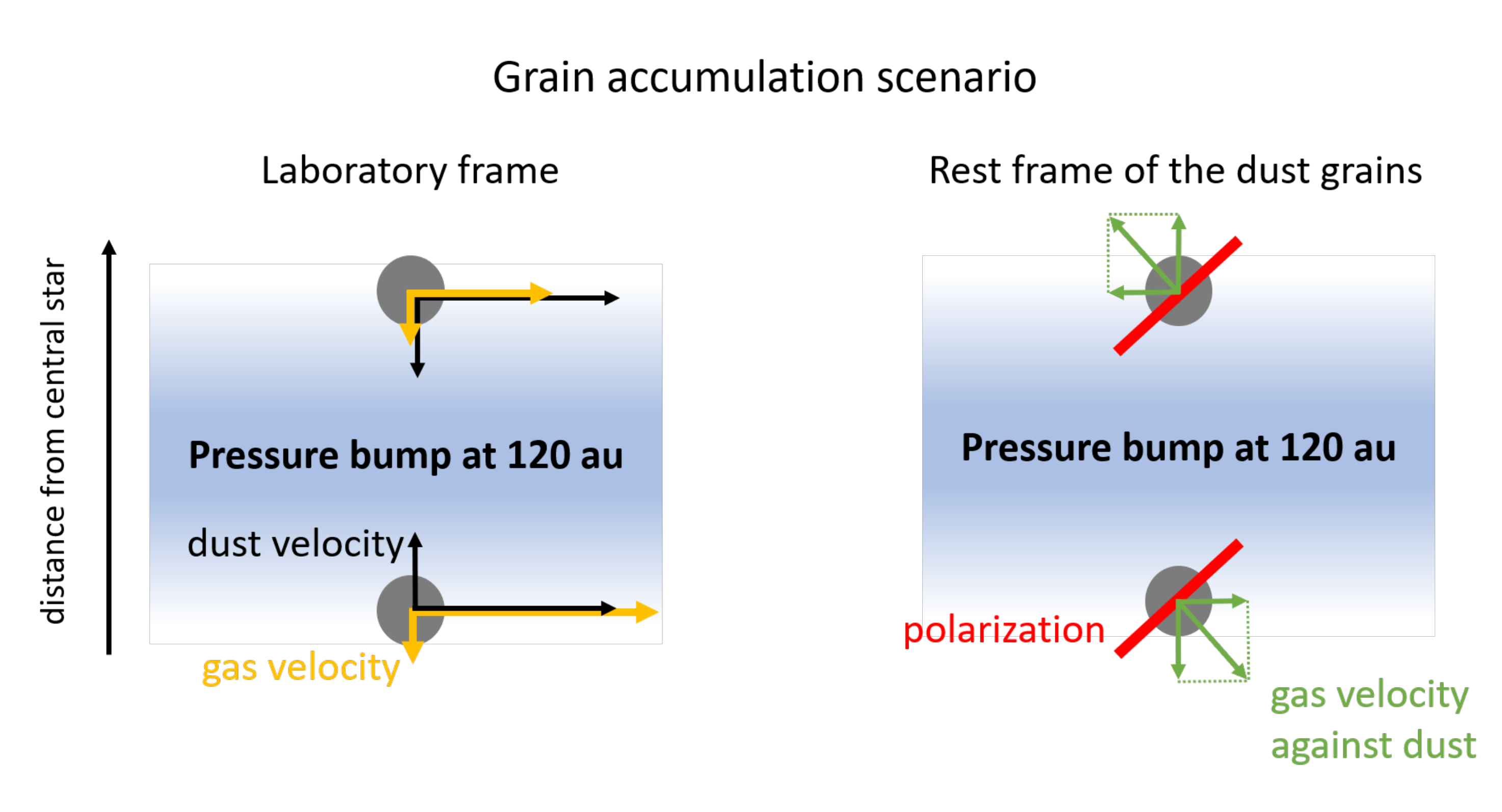}
\caption{Schematic illustration of the grain accumulation scenario which can reproduce the inclined polarization orientations. The blue region represents the gas pressure bump near the 120 au ring. The left picture shows grain (black) and gas (orange) velocity fields in the laboratory frame. The right picture shows the radial, azimuthal and synthetic gas velocity against dust grains with green arrows. The resultant polarization vectors are also presented with the red lines.}
\label{fig:drifting_scenario}
\end{figure}

Figure \ref{fig:drifting_scenario} illustrates a possible scenario, where the dust grains radially drift inside and outside of the local pressure maxima. 
To find the resultant polarization pattern at the pressure maxima, we discuss the velocity vectors of gas and dust grains.
In a laboratory frame, the gas rotates with sub-Keplerian outside the bump due to the negative pressure gradient while it rotates with super-Keplerian inside the bump due to the positive pressure gradient.
Instead, the dust grains rotates almost with the Keplerian speed.
As a result, the dust grains in both sides of the bump drift to the pressure maximum.
Now, we see the velocity fields on the rest frame of the dust grains.
The radial and azimuthal components of the relative gas velocity, $\delta$$v_{r}$ and $\delta$$v_{\phi}$ become comparable. 
Because the polarization vectors are perpendicular to the direction of the gas velocity on the rest frame of the dust grains in the helicity-induced alignment model, this leads to the spiral-like polarization patterns.

If the Stokes number is smaller than unity, dust grains would satisfy $\delta$$v_{r} > \delta$$v_{\phi}$.
This leads the polarization pattern almost in the azimuthal direction, but slightly deviates to the clockwise or counterclockwise directions depending on the rotation direction: if the disk rotates in the counter-clockwise direction, the polarization vectors are slightly deviates in clockwise direction (see \citet{kataoka2019}).
The direction of the disk rotation has been revealed to be clockwise with previous CO isotopologue observations \citep{andrews2009, guzman2018, favre2019}.
Therefore, the expected direction of the deviation is in the counterclockwise direction, or the plus sign in the histogram as shown in Figure \ref{fig:outer_histogram} while the figure shows the opposite results (see the discussion below). 

This scenario that produces the spiral-like pattern is partly supported by the previous CO isotopologue observation, where local enhancement of the gas surface density was discovered near the outermost dust ring \citep{favre2019}. This suggests the presence of the local gas pressure maximum which coincident with the dust ring, implying that we see the drifting grains in the pressure bump with the polarization.

We here note that the grain diffusion from the ring also reproduces the same polarization pattern as the accumulation model.
We do not discuss the diffusion but accumulation because it is more likely to occur at a ring.

The common scenario to make the gas pressure maxima are, for example, a planet \citep{kanagawa2015, kanagawa2016, bae2017, dong2017, huang2018, fedele2018, favre2019} and some MHD effects \citep{johansen2009, bai2014, simon2014}.
We note that the grain accumulation can also be triggered by secular gravitational instability \citep{tominaga2018}. 
This model predicts that the gas also be accumulated and thus the gas velocity on the dust grains is slightly different from the models above. 
While we do not discuss the detailed comparison, the presented polarization may distinguish these physics by revealing the relative velocity between gas and dust grains.

The ratio of the radial and azimuthal components of gas velocity against the dust grains can be derived by assuming the models as $\delta$$v_{\phi}$/$\delta$$v_{r}$ = tan($\Delta$$\theta_{ellipse}$) = 0.08 $\pm$ 0.03.
Both of the models predict the same directions of the angular deviations, which are determined by which clockwise or counterclockwise the disk rotates.
When the disk rotates clockwise, as shown in Figure \ref{fig:drifting_scenario} shows, the orientations are inclined to the west directions, which correspond to positive angular deviations.

However, there are several questions for the interpretation.
One is that the observed pattern is more consistent with the elliptical pattern, which is inconsistent with the theoretical expectations of the helicity-induced alignment model \citep{kataoka2019}.
Moreover, both of the models predict the positive angular deviations whereas the observed orientations show negative angular deviations (i.e. the theory predict 90$^\circ$ flipped orientations against the observed orientations). Thus, the directions of the orientations cannot be reproduced with the combination of the helicity-induced alignment model and the grain accumulation model as long as the disk rotates clockwise.

Since we find that the helicity-induced alignment model does not perfectly explain the observations, we also consider another mechanical alignment, the Gold mechanism although it probably does not occur in disks since gas velocity against dust grain is subsonic.
As explained above, the polarization orientations are parallel to the gas flow against the dust grains for the Gold mechanism, leading to the elliptical polarization pattern \citep{yang2019}.
This expectation is consistent with the fact that the observed pattern is more consistent with the elliptical pattern. The Stokes number should be much larger than unity (St $\gtrsim$ 10) so that $\delta$$v_{\phi}$ is much larger than $\delta$$v_{r}$ and the azimuthal pattern arises.
If it is the case in the disk, the polarization orientations are inclined with the same directions of the observed polarization. However, another question is that the thermal emission from the dust grains with such large Stokes number is not efficient because the emissivity of the dust grains is inversely proportional to the grain size. Therefore, even though Gold mechanism is considered, it is uncertain whether the observed orientations can be explained by the possible grain dynamics.

We summarize that the observed spiral-like pattern can be reproduced only with the mechanical alignment model. However, both Gold mechanisms and helicity-induced alignment models have some difficulties, which prevent us from naturally interpreting the orientations by assuming the grain dynamics.
This poses the possibility that we misunderstand somewhere in the alignment processes and/or grain dynamics in the protoplanetary disks. 

We also discuss the observed polarization fraction. We detect $\sim$$1.0$\% polarization fractions in the outermost regions. 
Moreover, we also find the asymmetry that the polarization fraction in the north region, which is farther to us reaches the larger value with 1$\sigma$ (Figure \ref{fig:polarization_fraction}). No theoretical expectations for the polarization fractions for the mechanical alignment model have been established so far, and thus it is unclear what determines the values and spatial distributions of the polarization fractions. 
The near/far side asymmetry is thought to be observed in the self-scattering and optically thick disks. 
\citet{yang2017} predicted a spatial shift of the polarized intensity peak and near/far side asymmetry of the polarization fractions, both of which are caused by geometrical effects such as disk flaring.
However, the AS 209 disk was revealed not to be optically thick at 870 $\micron$ \citep{perez2012, tazzari2016}, casting a question for applying the model to the disk.
In fact, the center shift, which was predicted by \citet{yang2017}, is not observed in the inner region of the AS 209 disk, where the scattering-induced polarization is observed. 
Therefore, the optical depth effects are unlikely to reproduce the observed profile and the origin of the profile is uncertain. This will be investigated in future works.

\section{Conclusions}
We have presented 870 $\mu$m polarization observation toward the Class I\hspace{-.1em}I protoplanetary disk around AS 209.
Our main findings can be summarized as follows:
\begin{enumerate}
\item We found the spatial segregation of the polarization patterns and fractions between inner and outer regions. We detected $\sim$0.2\% polarization in the inner 0.\arcsec5 regions. The polarization orientations in the region are parallel to the minor axis. In the outer region, we detected $\sim$1.0\% polarization and found that the polarization orientations are consistent with the elliptical pattern but with the angular deviation of $\Delta$$\theta_{ellipse}$ $\sim$ 4\fdg5 $\pm$ 1\fdg6. 
\item The polarization pattern in the inner region is consistent with the self-scattering model. The low polarization fraction ($\sim$0.2\%) compared to the expected value when the dust grains population has a single power law with $a_{max}$ $\sim$ $\lambda$/2$\pi$ $\sim$ 140 $\mu$m can be explained by the following grain models.
One is that the grains population has a single power law with $a_{max}$, which is a few times larger or smaller than 140 $\micron$. The other is that there is another grain population, which contributes to unpolarized emission but not to the polarized emission.
\item The spiral-like pattern in the outer region can be produced only with the mechanical alignment model. This polarized emission likely comes from the outermost ring at 120 au.
This suggests that the dust grains at the ring accumulate to the pressure maximum near 120 au.
\item We found that the combination of mechanical alignment and grain accumulation model can reproduce spiral-like pattern, but there are some inconsistencies.
The helicity-induced model predicts (1) circular polarization pattern and (2) positive angular deviations from the azimuthal directions, both of which are inconsistent with the observed pattern. The Gold mechanism can reproduce both of the elliptical pattern and the observed deviations from the azimuthal directions, but only in the case that the Stokes number is large, where the grain emissivity is inefficient in the millimeter-wave.
\vspace{-0.5mm}
\item No theoretical expectations for the polarization fraction for the mechanical alignment disks. Thus, it is yet clear whether the obtained values and spatial distribution of the polarization fractions are typical or not for mechanical alignment disk. This will be investigated in future works.
\end{enumerate}

For the detailed analysis of the polarization angles presented in this work, the large beam size effects are not negligible. 
Observations with higher spatial resolution would help to confirm the presence of the spiral-like pattern and understand the origin.

\acknowledgments
We gratefully appreciate Takashi Miyata, Takafumi Kamizuka, and Ryou Ohsawa for fruitful discussions for our study. This work is supported by JSPS KAKENHI Numbers 18K13590, 19H05088, 19K03932, 18H05441 and 17H01103.
This paper makes use of the following ALMA data: ADS/JAO.ALMA$\#$2017.1.00124.S.
ALMA is a partnership of ESO (representing its member states), NSF (USA) and NINS (Japan), together with NRC (Canada), MOST and ASIAA (Taiwan), and KASI (Republic of Korea), in cooperation with the Republic of Chile. The Joint ALMA Observatory is operated by ESO, AUI/NRAO, and NAOJ.

Data analysis was partly conducted with common use data analysis computer system at the Astronomy Data Center, ADC, of the National Astronomical Observatory of Japan.

\vspace{5mm}
\facilities{ALMA}


\software{CASA v5.1.1; \citep{mcmullin2007}
          }


\vspace{30mm}
\appendix
\section{Statistical tests for examining the mechanical alignment model}
We conduct a simple statistical analysis of the polarization angles to examine (1) which circular or elliptical pattern is compatible with the observed pattern, and (2) whether systematic angular deviations exist or not.
\begin{figure}[h!]
\centering
\includegraphics[height=7cm]{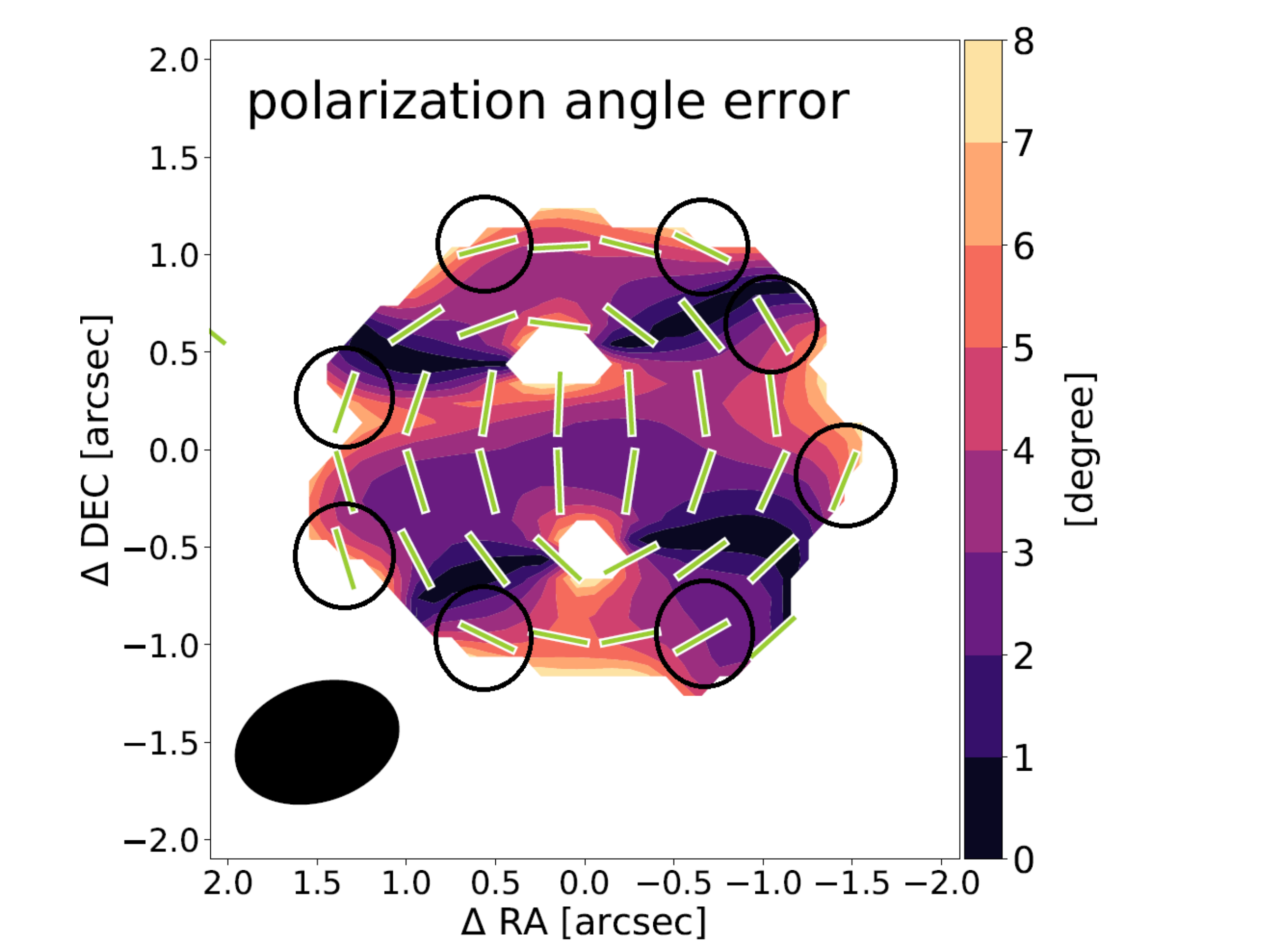}
\caption{The polarization angle error map and vectors. The vectors which are used for the statistical analysis are encirced in black.}
\label{fig:universe}
\end{figure}

For the analysis, each polarization angle should be independently sampled without the overlap of the synthesized beam.
Thus, as shown in Figure 7, we select the eight positions in the outer region, where the contamination from the central emission likely weak. Since the polarization angle error is obtained at the each position, the value of $\chi^2$ can be calculated with the following equation.
\begin{equation}
\chi^2 = \sum_{i = 1}^{8}\frac{\{{\theta_{obs, i} - (\theta_{circle/ellipse, i} + \Delta\theta_{circle/ellipse})}\}^2}{\sigma_{\theta, i}^2}
\end{equation}
where $\theta_{obs, i}$, $\theta_{circle/ellipse, i}$ and $\sigma_{\theta, i}$ are the observed polarization angle, the position angle of the circular/elliptical tangent, and the polarization angle error at $i$th position, respectively. $\Delta$$\theta_i$ is weighted mean of the angular deviation from circular/elliptical tangents at $i$th position, which can be estimated with a following equation.
\begin{equation}
\Delta\theta_{circle/ellipse} = \sum_{i = 1}^{8}\frac{(\theta_{obs, i} - \theta_{circle/ellipse, i})^2/\sigma_{\theta, i}^2}{1/\sigma_{\theta, i}^2}
\end{equation}
The uncertainty of $\Delta\theta_{circle/ellipse}$, $\delta\theta$ can also be estimated with 
\begin{equation}
\delta\theta = \sum_{i = 1}^{8}\frac{1}{\sigma_{\theta, i}^2}
\end{equation}
First, with the equation (A2) and (A3), we calculate the weighted mean of the angular deviation from circular/ellipse tangents and their uncertainties. Then, we substitute the values to equation (A1) for the circular and elliptical cases.
The equation (A1) yields $\Delta\theta_{circle}$ = $-$3\fdg8$^\circ$ $\pm$ 1\fdg6 and $\Delta\theta_{ellipse}$ = $-$4\fdg5$^\circ$ $\pm$ 1\fdg6, each of which corresponds to $\chi^2$ = 25 and 10, respectively.
Thus, we conclude that the elliptical pattern is preferred rather than the circular pattern. This is the answer for (1) described above. At the same time, the mean value of the angular deviations from the elliptical tangents is $\Delta\theta_{ellipse}$ = $-$4\fdg5 $\pm$ 1\fdg6, which significantly deviates from 0.
Therefore, we also conclude that there is the systematic angular deviation from the elliptical tangents. This is the answer for (2).



\end{document}